\documentclass[12pt,letterpaper]{JHEP3}
\usepackage{graphics}
\usepackage{epsfig}

\title{Note on the thermal history of decoupled massive particles}
\author{Hongbao Zhang\\
Perimeter
Institute for Theoretical Physics, Waterloo, Ontario, N2L 2Y5,
Canada\\
Department of Astronomy, Beijing Normal University, Beijing,
100875, China\\
\email{hzhang@perimeterinstitute.ca}}

\abstract{This note provides an alternative approach to the momentum decay and thermal evolution of decoupled massive particles. Although the ingredients in our results
have been addressed in Ref.\cite{Weinberg}, the strategies employed here are simpler, and the results obtained here are more general.}

\begin{document}
\section{Introduction}
As is well known, for the freely traveling massless particle like
photon in an expanding FLRW universe, the frequency or energy will
vary inversely proportional to the scale factor, which implies that
the number density of massless particles still keeps its thermal
spectrum form with a redshifted effective temperature although these
particles went out of the thermal equilibrium into the free
expansion as time passed. This is the physical foundation for the
cosmic microwave radiation background currently observed by us. Now
a natural question arises, namely, does the above fact also apply to
the massive particle? Not only does this question possess a
theoretical interest by itself, but also acquires a practical
implication in cosmology since neutrinos and antineutrinos are
believed to be massive. However, to my best knowledge, this issue
has not been addressed in literatures except in Weinberg's cosmology
book published recently\cite{Weinberg}.

The purpose of this note is to provide an alternative approach to
this issue. The strategies employed here are simpler, but the
results obtained here are more general. Notations and conventions
follow Ref.\cite{Wald}.
\section{Momentum Decay}
In general curved spacetime, a particle of mass $m$ freely travels
along the timelike geodesic $\eta(\tau)$ with $\tau$ the proper
time, which means that $U^a=(\frac{\partial}{\partial\tau})^a$ gives
the geodesic equation
\begin{equation}
U^a\nabla_aU^b=0
\end{equation}
with $U^aU_a=-1$. Assume there to be a family of observers $Z^a$
along the geodesic, then we have
\begin{equation}
\frac{dE}{d\tau}=U^a\nabla_a(-mU^bZ_b)=-mU^aU^b\nabla_aZ_b,\label{general}
\end{equation}
where $E=-mU^bZ_b$ is the energy of massive particle measured by the
observers.

Now for the expanding FLRW metric
\begin{equation}
ds^2=-dt^2+a^2(t)[\frac{dr^2}{1-Kr^2}+r^2(d\theta^2+\sin^2\theta d\varphi^2)]
\end{equation}
with $K=1,0,-1$ for closed, flat, and open universes respectively,
if the observers are chosen to be the isotropic ones as usual, i.e.,
$Z^a=(\frac{\partial}{\partial t})^a$, we have
\begin{equation}
\nabla_aZ_b=\frac{\dot{a}}{a}h_{ab},\label{special}
\end{equation}
where the dot denotes the derivative with respect to the time $t$,
$h_{ab}$ is the induced metric on the surface of constant $t$, given
by $h_{ab}=g_{ab}+(dt)_a(dt)_b$. Plugging Eq.(\ref{special}) into
Eq.(\ref{general}), we obtain
\begin{equation}
\frac{dE}{d\tau}=-m\frac{\dot{a}}{a}U^aU^bh_{ab}=-m\frac{\dot{a}}{a}[-1+(U^aZ_a)^2]=-\frac{E^2-m^2}{m}\frac{\dot{a}}{a},
\end{equation}
which implies
\begin{equation}
-\frac{da}{a}=m\frac{dt}{d\tau}\frac{dE}{E^2-m^2}=\frac{EdE}{E^2-m^2}=\frac{1}{2}\frac{d(E^2-m^2)}{E^2-m^2}=\frac{dp}{p},
\end{equation}
where $p=\sqrt{E^2-m^2}$ is the magnitude of momentum of massive
particle measured by the isotropic observers. Whence we know that
for a freely traveling massive particle in an expanding FLRW
universe, it is its momentum rather than energy that goes
like\footnote{Of course, the momentum of a massless particle shares
the same behavior since its momentum equals energy.}
\begin{equation}
p\propto\frac{1}{a}.\label{decay}
\end{equation}
It is noteworthy that this result is also obtained in
Ref.\cite{Weinberg}, where, however, the method employed seems
somewhat complicated, and some approximations are also made.

\section{Thermal Evolution}
Let us assume that during the evolution of our universe, there
exists a last scattering surface at the time $t_L$ when some kinds
of massive particles such as neutrinos and antineutrinos suddenly
went from being in thermal equilibrium to a decoupled expansion.
Then according to Eq.(\ref{decay}) the massive particle that has
momentum $p$ at a later time $t$ would have had momentum
$p_L=p\frac{a(t)}{a(t_L)}$ at the time $t_L$. So the number density
of massive particles at the time t with momentum between $p$ and
$p+dp$ would be
\begin{eqnarray}
n(p,t)dp&=&(\frac{a(t_L)}{a(t)})^3n(p_L,t_L)d(p_L)\nonumber\\
&=&(\frac{a(t_L)}{a(t)})^3\frac{4\pi gp_L^2dp_L}{(2\pi\hbar)^3}\frac{1}{\exp{[(\sqrt{p_L^2+m^2}-\mu_d)/kT_d]}\pm1}\nonumber\\
&=&\frac{4\pi gp^2dp}{(2\pi\hbar)^3}\frac{1}{\exp{[(\sqrt{p_L^2+m^2}-\mu_d)/kT_d]}\pm1}\nonumber\\
&=&\frac{4\pi gp^2dp}{(2\pi\hbar)^3}\frac{1}{\exp{[(\sqrt{p^2+m_e^2}-\mu_e)/kT_e]}\pm1}.
\end{eqnarray}
Here the factor $(\frac{a(t_L)}{a(t)})^3$ in the first step arises
from the dilution of particles due to the cosmic expansion. The
Fermi-Dirac and Bose-Einstein distributions are employed in the
second step, where $g$ is the number of spin states of the particle
and antiparticles, $\mu_d$ and $T_d$ denote the chemical potential
and temperature in thermal equilibrium at the last scattering
surface, respectively, and the sigh is $+$ for fermions and $-$ for
bosons. We introduce the effective mass, chemical potential, and
temperature in the last step, i.e.,
$m_e=m\frac{a(t_L)}{a(t)},\mu_e=\mu_d\frac{a(t_L)}{a(t)},T_e=T_d\frac{a(t_L)}{a(t)}$.
Therefore the form of the Fermi-Dirac and Bose-Einstein
distributions are preserved for the thermal evolution of decoupled
massive particle, with the effective mass, chemical potential, and
temperature varying inversely proportional to the scale factor $a$
at the same time, which implies that the ratios among the effective
mass, chemical potential, and temperature remain constant, just as
before decoupling.

Note that just by taking the mass to be zero the above argument
obviously reduces to the massless case, where the result is also
obtained in Ref.\cite{Weinberg} by the thermodynamic method rather
than the simpler dynamical picture employed here. In addition,
although the spectrum has still kept the form of the Fermi-Dirac and
Bose-Einstein distributions since decoupling, it is not the thermal
spectrum with the effective temperature and chemical potential since
the effective mass is not equal to the static mass. The unique
exception is the massless case.

\acknowledgments The author would like to give much gratitude to
Steven Weinberg for his helpful correspondence on this work. The
author was supported in part by the Government of China through
CSC(no.2007102530). This research was supported by Perimeter
Institute for Theoretical Physics. Research at Perimeter Institute
is supported by the Government of Canada through IC and by the
Province of Ontario through MRI.

\end{document}